%% file: paper.tex
\documentclass[superscriptaddress,twocolumn]{revtex4}
\usepackage{amsmath,lscape,epsfig}

\usepackage[latin1]{inputenc}

\def\beq{\begin{equation}}
\def\eeq{\end{equation}}
\def\beqa{\begin{eqnarray}}
\def\eeqa{\end{eqnarray}}
\def\ban{\begin{eqnarray*}}
\def\ean{\end{eqnarray*}}
\def\bi{\begin{itemize}}
\def\ei{\end{itemize}}

\def\d{\mbox{d}}

\begin{document}

\title{Warm pasta phase in the Thomas-Fermi approximation}

\author{Sidney S. Avancini}
\affiliation{Depto de F\'{\i}sica - CFM - Universidade Federal de Santa
Catarina  Florian\'opolis - SC - CP. 476 - CEP 88.040 - 900 - Brazil}
\author{Silvia Chiacchiera}
\affiliation{Centro de F\'{\i}sica Computacional - Department of Physics -
University of Coimbra, P3004 - 516, Coimbra, Portugal}
\author{D\'ebora P. Menezes}
\affiliation{Depto de F\'{\i}sica - CFM - Universidade Federal de Santa
Catarina  Florian\'opolis - SC - CP. 476 - CEP 88.040 - 900 - Brazil}
\author{Constan\c ca Provid\^encia}
\affiliation{Centro de F\'{\i}sica Computacional - Department of Physics -
University of Coimbra, P3004 - 516, Coimbra, Portugal}

\begin{abstract}
In the present paper, the pasta phase is studied at finite temperatures
within a Thomas-Fermi (TF) approach. Relativistic mean field  models, both with
constant and density-dependent couplings, are used to describe this
frustrated system. We compare the present results with previous ones obtained
within a phase-coexistence description and conclude that
the TF approximation gives rise to a richer inner pasta phase structure
and the homogeneous matter appears at higher densities. 
Finally, the transition density calculated within TF is compared with
the results for this quantity obtained with other methods.

\end{abstract}
\maketitle

\vspace{0.50cm}
PACS number(s): {21.65.-f, 24.10.Jv, 26.60.-c, 95.30.Tg}
\vspace{0.50cm}

\section{Introduction}

A complete theoretical description of the processes involved in supernova
explosion and stellar evolution depends on the construction of an adequate
equation of state, able to describe matter ranging from very
low densities to few times saturation density and from zero temperature to
around 100 MeV. Protoneutron stars are believed to be the initial phase of the
remnants of a supernova explosion. They cool down to become a cold
neutrino-free neutron star by emitting neutrinos, which means that the
neutrino mean free path inside the star is an important quantity in the
understanding of the stellar evolution. The neutrino mean free path depends on
the reactions that may take place inside the star. Hence, the composition of
the star plays a definite role in its fate.

Protoneutron stars are believed to have a crust where a special matter known as
{\it pasta phase} is expected to be present.
The pasta phase is a frustrated system that arises in the
competition between the strong and the electromagnetic interactions
\cite{pethick,horo,maruyama,watanabe05,watanabe08}. The pasta phase appears
at densities of the order of 0.001 - 0.1 fm$^{-3}$ \cite{pasta1,watanabe08}
in neutral nuclear matter 
and in a smaller density range \cite{bao,pasta2}
in $\beta$-equilibrium stellar matter.
The basic shapes of these structures were named \cite{pethick}
as droplets (bubbles), rods (tubes) and slabs for three, two and one dimensions
respectively. The ground-state configuration is the one that minimizes the
free energy, i.e., the pasta phase is the ground-state configuration if its
free energy per particle is lower than the one corresponding to the homogeneous
phase at the same density.

In previous works \cite{pasta1,pasta2} we have studied the
existence of the pasta phase at zero and finite temperature within
different parametrizations of the relativistic non-linear Walecka model
(NLWM) and of the density-dependent hadronic model. In the present work we will 
consider the NLWM parametrization NL3 \cite{nl3} and the density-dependent
hadronic model TW \cite{tw}.
These two models do not satisfy some of the constraints on the high density behaviour of the 
equation of state (EOS) discussed in \cite{klaehn06}.
However,  both models  describe reasonably well the groundstate properties of stable and unstable nuclei, and, therefore, are adequate to study the EOS at  subsaturation densities. Moreover,  we are interested in studying the effect of  the density dependence of the symmetry energy on the pasta phase. Therefore, we have chosen a model (NL3) with a very large symmetry energy   slope at saturation, above the limit imposed by isospin diffusion \cite{chen05}, and  another one with a slope close to the lower limit defined by the same experiments.
In both works \cite{pasta1,pasta2}
two different methods were used: the coexisting phases
(CP), both at zero and finite temperature, and the Thomas-Fermi (TF)
approximation at zero temperature only. It was found that if 
$\beta$-equilibrium is imposed the pasta phase does not appear in a CP
calculation for the same surface energy parametrization used for fixed proton
fractions. This indicates the necessity to use a good parametrization for the
surface energy which is temperature, proton fraction and geometry dependent,
as also stressed in \cite{gogelein,dalen}. The specific problem of an
appropriate parametrization for the surface energy was tackled in
\cite{pasta_alpha}.

In \cite{opacity} it is found that the diffusion coefficients, related to
neutrino opacities in dense matter, are significantly altered by the
presence of the nuclear pasta in stellar matter. 
These differences in neutrino opacities
certainly influence the Kelvin-Helmholtz phase of protoneutron stars as well
as supernova explosion simulations.

In the present paper we use the Thomas-Fermi approximation to obtain the
pasta phase for two parametrizations, namely NL3 and TW,
and compare the results with the ones obtained with the more naive 
coexisting-phases method at finite temperature.
 The CP method used here is improved
with respect to references \cite{pasta1,pasta2} because it takes into account
a better description of the surface energy, as in \cite{pasta_alpha}.
We check the differences in phase transition from pasta to homogeneous matter
and see how different the internal structures of the pasta are. Some preliminary results for the pasta phase obtained at finite temperature within the Thomas
Fermi approximation have been published in \cite{avancini10,maruyama10}.

It was shown in \cite{link} that the pressure and density at the
inner boundary of the crust 
(transition pressure and transition density)
define the mass and moment of
inertia of the crust. This establishes a relation between 
the equation of state (EoS) and compact-stars observables. 
Several methods have been used to calculate this density 
for $\beta$-equilibrium cold matter: a local equilibrium approximation 
\cite{pethick95,xu09}, the thermodynamical spinodal method 
\cite{oyamatsu07,pasta1,pasta2,xu09}, and 
the dynamical spinodal method used in \cite{ducoin08,pasta1,pasta2} 
which predicts a transition density very close to the Thomas-Fermi result. 
It is expected that the transition density lies in the metastable region  
between the binodal surface and the dynamical spinodal surface. 
We will compare the different predictions for the transition density 
at several temperatures and isospin asymmetries obtained from the 
thermodynamical spinodal, the dynamical spinodal, the binodal surface 
and the TF and CP calculations. This will allow us to make an estimation 
of the error done within each method.

In section \ref{sec:formalism}
the basic expressions for the non-linear Walecka model are
outlined. In section \ref{sec:pastatf}
we briefly describe the TF approximation
at finite temperature  and in section \ref{sec:pastacp} 
the CP method is
reproduced. In section \ref{sec:results} the results are shown and discussed and in section 
\ref{sec:conclusions}
the final conclusions are drawn.

\section{The Formalism}\label{sec:formalism}

We consider a system of protons and neutrons with mass $M$
interacting with and through an isoscalar-scalar field $\phi$ with mass
$m_s$,  an isoscalar-vector field $V^{\mu}$ with mass
$m_v$, an isovector-vector field  $\mathbf b^{\mu}$ with mass
$m_\rho$.  We also include a system of electrons with mass $m_e$.
The Lagrangian density reads:

\begin{equation}
\mathcal{L}=\sum_{i=p,n}\mathcal{L}_{i}+\mathcal{L}_e\mathcal{\,+L}_{{\sigma }}%
\mathcal{+L}_{{\omega }}\mathcal{+L}_{{\rho }}\mathcal{+L}_{{\gamma }},
\label{lagdelta}
\end{equation}
where the nucleon Lagrangian reads
\begin{equation}
\mathcal{L}_{i}=\bar{\psi}_{i}\left[ \gamma _{\mu }iD^{\mu }-M^{*}\right]
\psi _{i}  \label{lagnucl},
\end{equation}
with
\begin{eqnarray}
iD^{\mu } &=&i\partial ^{\mu }-\Gamma_{v}V^{\mu }-\frac{\Gamma_{\rho }}{2}{\boldsymbol{\tau}}%
\cdot \mathbf{b}^{\mu } - e \frac{1+\tau_3}{2}A^{\mu}, \label{Dmu} \\
M^{*} &=&M-\Gamma_{s}\phi.
\label{Mstar}
\end{eqnarray}
The electron Lagrangian density is given by
\begin{equation}
\mathcal{L}_e=\bar \psi_e\left[\gamma_\mu\left(i\partial^{\mu}+ e A^{\mu}\right)
-m_e\right]\psi_e,
\label{lage}
\end{equation}
and the meson Lagrangian densities are
\begin{eqnarray*}
\mathcal{L}_{{\sigma }} &=&+\frac{1}{2}\left( \partial _{\mu }\phi \partial %
^{\mu }\phi -m_{s}^{2}\phi ^{2}-\frac{1}{3}\kappa \phi ^{3}-\frac{1}{12}%
\lambda \phi ^{4}\right)  \\
\mathcal{L}_{{\omega }} &=&\frac{1}{2} \left(-\frac{1}{2} \Omega _{\mu \nu }
\Omega ^{\mu \nu }+ m_{v}^{2}V_{\mu }V^{\mu } \right) \\
\mathcal{L}_{{\rho }} &=&\frac{1}{2} \left(-\frac{1}{2}
\mathbf{B}_{\mu \nu }\cdot \mathbf{B}^{\mu
\nu }+ m_{\rho }^{2}\mathbf{b}_{\mu }\cdot \mathbf{b}^{\mu } \right)\\
\mathcal{L}_{{\gamma }} &=&-\frac{1}{4}F _{\mu \nu }F^{\mu
  \nu },
\end{eqnarray*}
where $\Omega _{\mu \nu }=\partial _{\mu }V_{\nu }-\partial _{\nu }V_{\mu }$
, $\mathbf{B}_{\mu \nu }=\partial _{\mu }\mathbf{b}_{\nu }-\partial _{\nu }\mathbf{b}%
_{\mu }-\Gamma_{\rho }(\mathbf{b}_{\mu }\times \mathbf{b}_{\nu })$
and $F_{\mu \nu }=\partial _{\mu }A_{\nu }-\partial _{\nu }A_{\mu }$.
The  parameters of the models  are: the nucleon mass $M=939$ MeV,
the coupling parameters $\Gamma_s$, $\Gamma_v$, $\Gamma_{\rho}$ of the mesons to
the nucleons, the electron mass $m_e$ and the electromagnetic coupling constant
$e=\sqrt{4 \pi/137}$.
In the above Lagrangian density $\boldsymbol {\tau}$ is
the isospin operator. When the TW density dependent model \cite{tw} is used,
the non-linear
terms are not present and hence $\kappa=\lambda=0$ and the density dependent
parameters are chosen as in \cite{tw,gaitanos,inst04}. When the NL3
parametrization is used, $\Gamma_i$ is replaced by $g_i$, where
$i=s,v,\rho$ as in \cite{nl3}. The bulk nuclear matter properties of the
models we use in the present paper are given in Table \ref{tab1}.We also
include in the table some properties at the thermodynamical spinodal surface:
$\rho_s$ is the upper border density at the spinodal surface  for symmetric
matter (it defines the density for which the incompressibility is zero),
$\rho_t$ and $P_t$ are, respectively, the density and the pressure at the
crossing between the cold $\beta$-equilibrium  equation of state and the
spinodal surface. They give a rough estimate of the density and pressure at
the crust-core transition \cite{pasta1,pasta2}.
The surface-tension coefficient $\sigma$ (see section \ref{sec:pastacp})
is also given for $T=0$ and two proton fractions, $Y_p=0.5$ and $Y_p=0.3$.

\begin{table}[h]
\caption{ Symmetric nuclear matter properties at the saturation density and 
at the spinodal surface. The surface tension $\sigma$, the  saturation 
density $\rho_0$ and the corresponding binding energy $B/A$  are
also given for the proton fraction  $Y_p=0.3$.}
\label{tab1}
\begin{center}
\begin{tabular}{lccccccccc}
\hline
&  NL3 & TW \\
&   \cite{nl3} & \cite{tw} \\
\hline
\hline
$\rho_0$ (fm$^{-3}$) & 0.148  & 0.153 \\
$B/A$ (MeV) & 16.3  & 16.3 \\
$\rho_0$ ($Y_p=0.3$) (fm$^{-3}$) & 0.118  & 0.135 \\
$B/A(Y_p=0.3)$ (MeV) & 10.9  & 11.2 \\
$K$ (MeV) & 272  & 240 \\
${\cal E}_{sym.}$ (MeV)  & 37.4  & 32.8 \\
$M^*/M$ & 0.60 & 0.55 \\
$L$ (MeV) & 118.3 & 55.3 \\
$K_{sym}$ (MeV) & 100.5 & -124.7 \\
$Q_0$ (MeV) & 203 & -540\\
$K_{\tau}$ (MeV) &-698 &-332\\
$\rho_s$ (fm$^{-3}$) &0.096 & 0.096\\
$\rho_t$ (fm$^{-3}$) & 0.065 & 0.085\\
$P_t$ (MeV/fm$^3$)& 0.396 & 0.455\\
$\sigma(Y_p=0.5)$ (MeV/fm$^2$)&  1.123 &       1.217\\     
$\sigma(Y_p=0.3)$ (MeV/fm$^2$)&  0.254 &       0.613\\     
\hline
\end{tabular}
\end{center}
\end{table}
We next give the basic expressions for the construction of the pasta phase
within the Thomas-Fermi calculation at finite temperature and of
the coexisting-phases method.

\section{The pasta phase}

\subsection{Thomas-Fermi approximation}\label{sec:pastatf}
In the present work we repeat the same numerical prescription given in
\cite{pasta1} where, within the Thomas-Fermi approximation of the
non-uniform {\it npe} matter, the fields are assumed to vary slowly so that
the baryons can be treated as moving in locally constant fields at each point.

 From a formal point of view the Thomas-Fermi
approximation can be considered as the
zeroth order term of a semiclassical expansion in the relativistic mean field theory
derived within the framework of the relativistic Wigner transform of operators
\cite{ring2,cente}. Here, we take a more straightforward way to obtain the finite temperature
semiclassical TF approximation
 based on the density functional formalism. We begin with the grand canonical potential density:
\begin{equation}
 {\omega } = \omega ( \{f_{i+}\},\{f_{i-}\},\phi_0,V_0,b_0  ) =
{\cal E}_t-T{\cal S}_t-\sum_{i=p,n,e}\mu_i \rho_i ~ ,~
\label{grand}
\end{equation}
above $\{f_{i+}\}$($\{f_{i-}\}$), i=p,n,e  stands for the  protons,
neutrons and electrons  positive(negative) energy distribution functions and
 ${\cal S}_t = {\cal S} +{\cal S}_e$ , ${\cal E}_t = {\cal E} +{\cal E}_e $ are
the total energy and entropy densities
respectively. The total energy has been defined in \cite{pasta1}  and for the entropy we take
the one-body entropy:
\begin{eqnarray}
 S_t&=&  -\sum_{i=n,p,e}\int d^3 r \int \frac{d^3 p}{4\pi^3} ~
\left\{ f_{i+}(\mathbf r, \mathbf p) \ln f_{i+} (\mathbf r, \mathbf p) \right.
\\
&&\left. +\left[1- f_{i+}(\mathbf r, \mathbf p)\right] \ln \left[1-f_{i+} (\mathbf r, \mathbf p)\right] +
( f_{i+} \leftrightarrow f_{i-} )
\right\}.\nonumber
\end{eqnarray}
where the  ground-state (equilibrium) distribution functions are
\begin{eqnarray}
f_{i\pm}(\mathbf r, \mathbf p)&=&\frac{1}{1+\exp
\left[{(\epsilon^\star(\mathbf r ,\mathbf p)\mp \nu_i(\mathbf r))/T}\right] }~ ,\, i=p,n\nonumber\\
\\
f_{e\pm}({\mathbf r},{\mathbf p})\,&=&\,\frac{1}{1+\exp[(\epsilon_e\mp
\nu_e(\mathbf r))/T]},
\end{eqnarray}
with $\epsilon^\star(\mathbf r, \mathbf p)=\sqrt{p^2+M^\star(\mathbf r)^2}$,
$M^\star (\mathbf r) = M- \Gamma_s \phi_0 (\mathbf r)$, and $\epsilon_e=\sqrt{p^2+m_e^2}$.
 $\nu_e(\mathbf r)=\mu_e+eA_0(\mathbf r)$ is the electron effective chemical potential and
the nucleon effective chemical potential, $\nu_i,\, i=p,n$, is given by:
\begin{equation}
\nu_i=\mu_i - \Gamma_v V_0 - \frac{\Gamma_{\rho}}{2}~  \tau_{3 i}~ b_0 - \frac{e}{2}(1+\tau_{3 i})A_0
-{\Sigma^{R}_{0}} ,
\end{equation}
where the rearrangement term is given by \cite{gaitanos}:
\begin{equation}
\Sigma^{R}_{0}(\mathbf r)=
\frac{\partial \Gamma_{v}}{\partial \rho} \rho(\mathbf r) V_0(\mathbf r) +
\frac{\partial \Gamma_{\rho}}{\partial \rho} \rho_3(\mathbf r) ~
\frac{b_0(\mathbf r)}{2} -
\frac{\partial \Gamma_s}{\partial \rho} \rho_{s}(\mathbf r) \phi_0(\mathbf r).
\label{rear}
\end{equation}

 The equations of motion for the meson fields (see \cite{pasta1})  follow from the variational 
conditions:
\begin{equation}
\frac{\delta}{\delta \phi_0(\mathbf r)} \Omega =
\frac{\delta}{\delta V_0(\mathbf r)} \Omega = \frac{\delta}{\delta b_0(\mathbf r)} \Omega =  0 ~ ,
\nonumber \label{meson}
\end{equation}
where
\begin{equation}
 \Omega = \int d^3 r \omega ( \{f_{i+}\},\{f_{i-}\},\phi_0,V_0,b_0  )  ~.
\label{grand1}
\end{equation}

The numerical algorithm for the description of the neutral $npe$ matter at finite temperature is
a generalization of the zero temperature case which
was discussed in detail in \cite{pasta1}.
The Poisson equation is always solved by using the appropriate Green
function according to the  spatial dimension of interest and the
Klein-Gordon equations are solved by expanding the meson fields in a harmonic
oscillator basis with one, two or three dimensions based on the method
proposed in \cite{ring}. The most important source of numerical problems are the Fermi integrals, hence,
we have used an accurate and fast algorithm given in ref.\cite{aparicio} for their calculations.

\subsection{Coexisting phases}\label{sec:pastacp}

In \cite{pasta1} a complete description of the coexisting-phases
method applied to different parametrizations of the NLWM is given.
In the description of the equations of state
of a system, the required quantities are the baryonic density $\rho$, the energy
density $\cal{E}$, the pressure $P$ and the free energy density $\cal F$, 
explicitly written in \cite{pasta1,pasta2}.

Two possibilities are discussed in the following: nuclear matter with fixed proton fraction
and $\beta$-equilibrium stellar matter. In the first case, electrons are 
related to the nuclear matter by the imposition of charge
neutrality in such a way that the electron density is equal to the proton
density. The equations
\begin{equation}
Y_p=\frac{\rho_p}{\rho}, \quad \rho_e=\rho_p
\end{equation}
hold. In the second case, the fractions of nucleons
and electrons are defined
by the conditions of chemical equilibrium and charge neutrality.
In this case, the enforced conditions are:
\begin{equation}
\mu_p=\mu_n-\mu_e, \quad \rho_e=\rho_p~,
\end{equation}
for neutrino-free matter
and
\begin{equation}
\mu_p=\mu_n-\mu_e + \mu_\nu, \quad Y_l=\frac{\rho_e+\rho_\nu}{\rho}, \quad
\rho_e=\rho_p~,
\end{equation}
if trapped neutrinos are considered.

As in \cite{pasta1,maruyama}, for a given total density $\rho$,
the pasta structures are built with different
geometrical forms in a background nucleon gas. This is achieved by calculating
from the Gibbs conditions the density and the proton fraction of the
pasta and of the background gas.

In building the pasta phase, the density of electrons is uniform.
The total pressure is given  by $P=P^I+P_e+P_\nu$ and the total energy density 
of the system is given by
\begin{equation}
{\cal E}= f {\cal E}^I + (1-f) {\cal E}^{II} + {\cal E}_e + {\cal E}_\nu
+ {\cal E}_{surf} + {\cal E}_{Coul},
\label{totener}
\end{equation}
where I and II label the high and low density phase respectively
and $f$ is the volume
fraction of phase I. Notice that matter with fixed proton fraction is
neutrino-free and hence the neutrino pressure and energy density are zero.
By minimizing the sum  ${\cal E}_{surf} + {\cal E}_{Coul}$ with respect
to the size of the droplet/bubble, rod/tube or slab we get
\cite{maruyama}
${\cal E}_{surf} = 2 {\cal E}_{Coul},$ and
\begin{equation}
{\cal E}_{Coul}=\frac{2 F}{4^{2/3}}(e^2 \pi \Phi)^{1/3}
\left(\sigma D (\rho_p^I-\rho_p^{II})\right)^{2/3},
\end{equation}
where $F=f$ for droplets and $F=1-f$ for bubbles,
 $\sigma$ is the surface energy coefficient,
$D$ is the dimension of the system and the geometric factor  $\Phi=
\left(\frac{2-D F^{1-2/D}}{D-2}+F \right) \frac{1}{D+2}, \, D=1,2,3.$
The surface coefficient $\sigma$, necessary in the 
above expressions, is given by \cite{marina,cpsig,dmcp}: 
\begin{equation}
\sigma = \int_{-\infty}^{\infty} dz \left( \left(\frac{d\phi_0}{dz}\right)^2 
-\left(\frac{dV_0}{dz}\right)^2 
- \left(\frac{db_0}{dz}\right)^2 \right)  ~, \label{sigcp}
\end{equation}
which is adequate for the parametrization of the dependence of $\sigma$ 
on the temperature. In the Appendix we show how to obtain this expression and 
the equivalence
between this expression and eq. (3.14) of reference \cite{centel}.

According to the calculations performed in \cite{pasta_alpha} the surface energy
in terms of the proton fraction varies considerably from the NL3 to the TW 
parametrization. In the first case it is always smaller. This quantity plays
an important role on the size and structures of the pasta phase discussed in the
present paper.

At this point it is worthy pointing out that the dependence of the
energy density on the electromagnetic and surface contributions is
commonly known as finite size effects. In \cite{maruyama07,vos2,yatsutake,
maru2010} it was shown that for a weak surface tension the EoS obtained in 
a mixed phase resembles the one obtained
with a Gibbs construction while, for a strong surface tension, the Maxwell
construction is reproduced. 
\begin{figure*}[htb]
\begin{center}
\begin{tabular}{cc}
\includegraphics[width=0.45\linewidth]{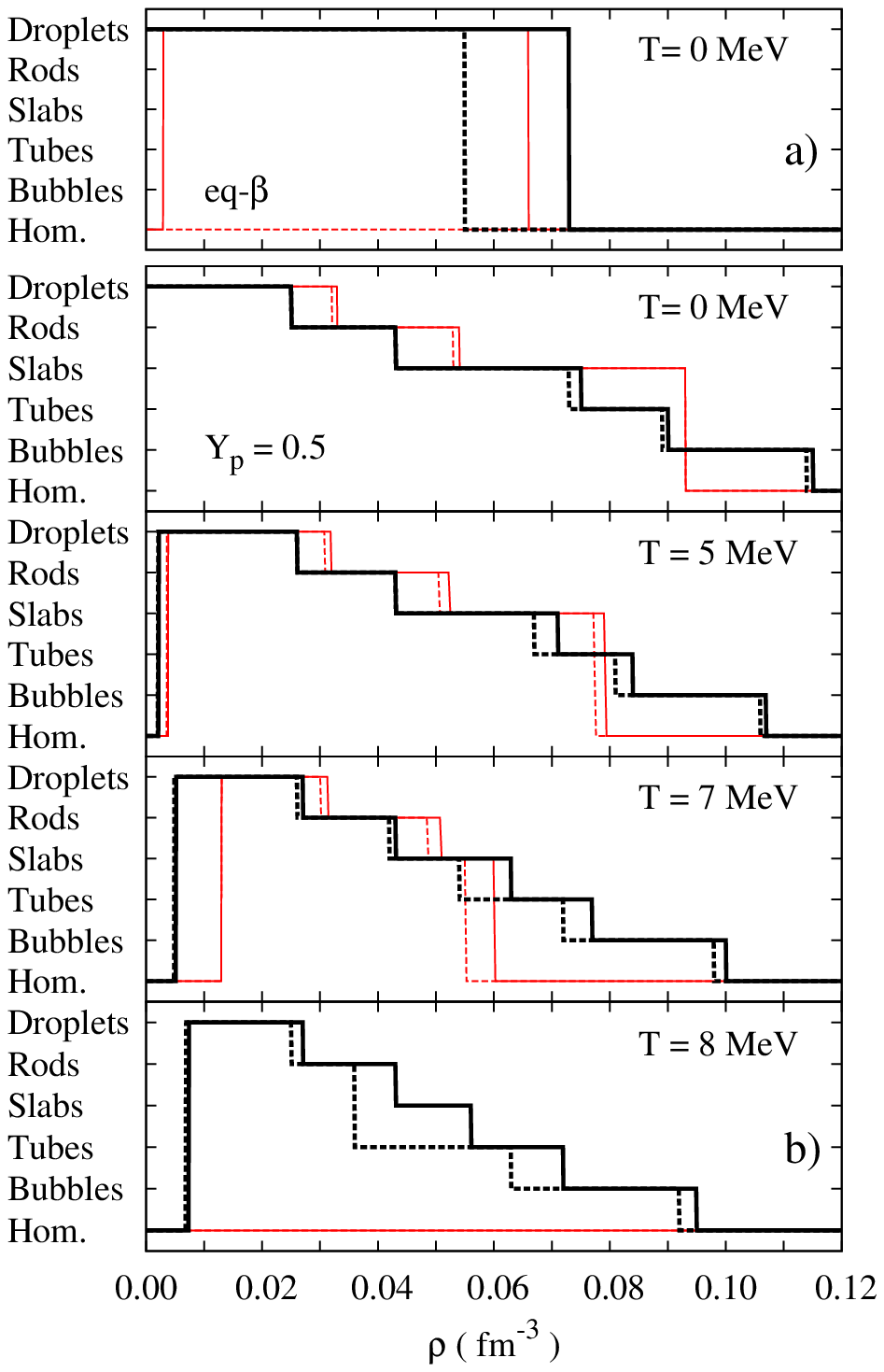} &
\includegraphics[width=0.45\linewidth]{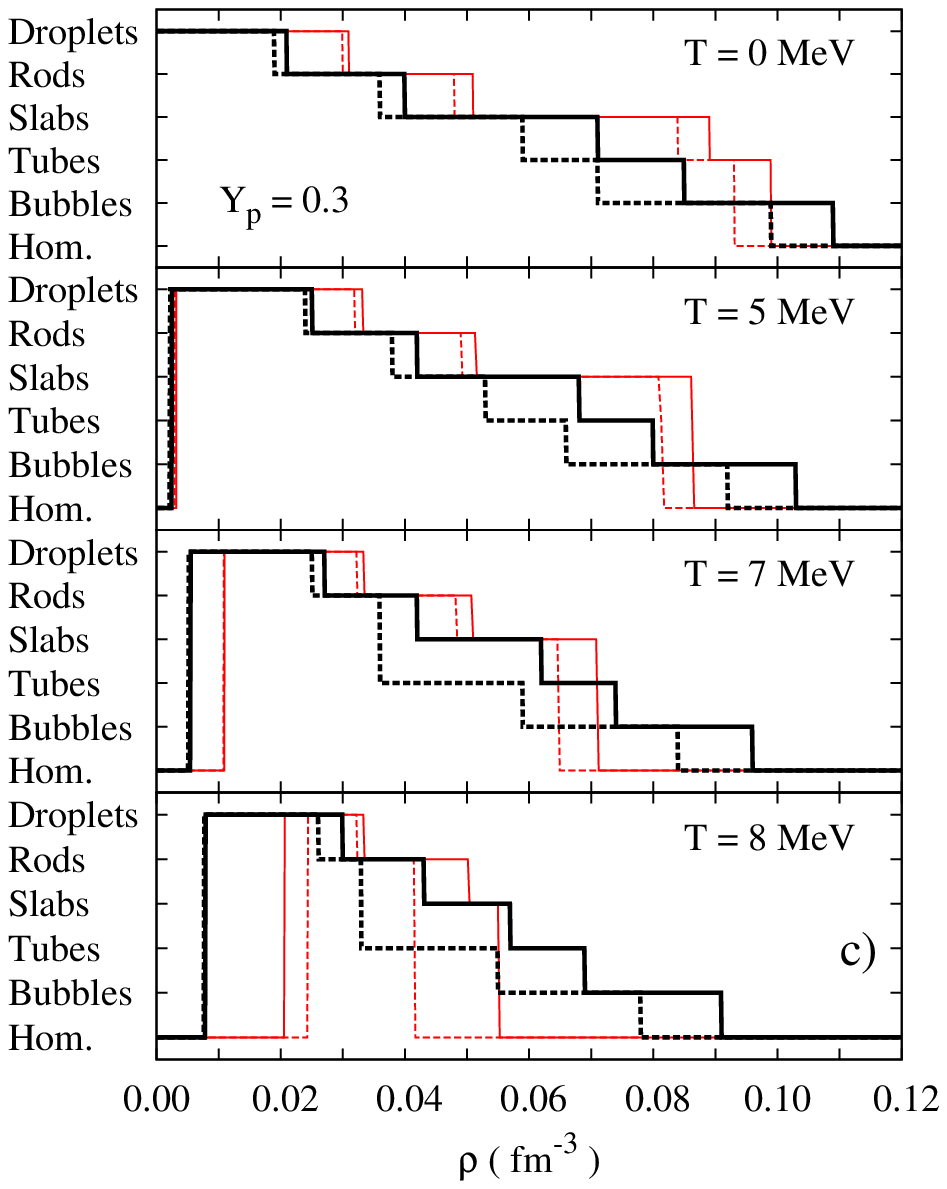} \\
\end{tabular}
\end{center}
\caption{Pasta phases: comparison between  NL3 (dashed line)  and TW (full line)
and the methods CP and TF for a) $\beta$-equilibrium cold stellar  matter; 
b) $Y_p=0.5$, c) $Y_p=0.3$. The thick lines refer to the Thomas-Fermi (TF) calculation and the thin red
lines to the coexisting-phases (CP) calculation.}
\label{fig2}
\end{figure*}

\section{Results and discussions}\label{sec:results}

In the present section we discuss the properties of the pasta phase obtained 
within the two methods described above. We consider several temperatures, 
$T=0, 5, 7, 8
$ MeV and two proton fractions $Y_p=0.5$ and $0.3$.
We only show results up to 8 MeV because it is not clear whether our framework 
model is realistic above this temperature
, since thermal fluctuations are not taken into account. 
The problem of the effect of  thermal fluctuations on the pasta structures 
has been studied by \cite{thermal1,thermal2} and it was 
shown that 
thermally induced displacements of the rod-like and
slab-like nuclei can melt the lattice structure when these displacement 
are 
larger than the space available between the cluster and the boundary of the 
Wigner-Seitz cell.  While for the rod like clusters $T=8$ MeV would still
be acceptable, 
for the slabs $T=8$ MeV could already be too large.
 
We perform most of our calculations for a fixed proton fraction, 
although matter in $\beta$-equilibrium with trapped neutrinos 
is known to be important.
In fact during  the first seconds of the protoneutron star, neutrinos are 
trapped. They start to diffuse out  10-15 s after the supernova explosion 
\cite{burrows86,prakash97}, when the fraction of
leptons decreases from a maximum value, $\sim$ 0.4,  constant through out the 
star, to smaller values.
Neutrino emission is essential to understand neutron star cooling
 \cite{yakovlev01,migdal90}. However, neutrinos couple weakly to nuclear matter,
therefore for the pasta calculation their presence mainly 
defines the proton fraction of the warm stellar matter.
 In \cite{ducoin08},
we have shown for a wide set of models, including NL3 and TW, that for the 
density  range of interest for the pasta phase, the proton fraction for the 
largest trapped fraction of neutrinos, which corresponds to a lepton fraction 
$Y_l\sim 0.4$, is approximately 0.3. Moreover, in a recent paper 
\cite{pasta_alpha}
we have studied the pasta phase in $\beta$-equilibrium stellar matter with 
trapped neutrinos ($Y_l=0.4$) within the more schematic coexisting phases 
approach and we have seen that the pasta phase structure and its 
extension is very similar to the one obtained for a fixed proton fraction 
of about 0.3.  

Since the inclusion of trapped 
neutrinos in the calculation does not bring much more information about the pasta phase 
itself (the main point of the present work), we have only 
included neutrinos in the present calculation for two cases 
(NL3 and TW, $T=5$ MeV and $Y_l=0.4$) in order to show the
similarity with the $Y_p=0.3$ results.

In Tables \ref{tabNL3} and \ref{tabTW} we display the onset densities for
each of the pasta structures and the homogeneous phase for the parametrizations
NL3 and TW respectively. The pressures identifying the starting ($P^1$) and 
ending ($P^2$) points of the pasta phase are also given. We investigate 
all possible structures for $T=0, 5,\, 7$ and 8 MeV and proton fraction equal
to 0.5 (symmetric matter) and $Y_p=0.3$. Both CP and TF approximations were
used.

From Table \ref{tabNL3} one can observe that TF always presents a richer inner
pasta phase structure as compared with the CP method. For $T=8$ MeV 
and symmetric matter, the pasta phase no longer exists if the CP method is 
used, but { almost}  all internal 
structures are present within the TF calculation. Generally, the homogeneous 
phase becomes the ground-state matter for densities much higher within the
TF approach than within the CP method. Just some exceptions were found for a 
complete structure (3D, 2D, 1D, 2D, 3D) with the TF calculation: $T=7$ MeV for 
$Y_p=0.3$ and
$T=8$ MeV for
$Y_p=0.3$ and $Y_p=0.5$, which lack the slab configuration. 
In all other cases, all possible
configurations were found. This is not true within a CP method, where many
structures were missing, giving rise to a much simpler pasta phase.

From Table \ref{tabTW} one sees that 
most of the conclusions drawn for the NL3 parametrization
hold true also for TW, as a richer inner pasta phase structure 
obtained within the TF approach than within the CP method. 
Once more the homogeneous 
phase becomes the ground-state matter for densities higher within the
TF approach than within the CP method. 
When the TW density dependent model is used, one can see that
all possible internal pasta configurations are always found with the TF 
approximation. This is not the case with the NL3 parametrization. 
Different pasta phase internal structures lead to different diffusion 
coefficients causing different neutrino mean free paths, as seen in 
\cite{opacity}. These differences in neutrino opacities may have consequences 
in calculations of protoneutron star evolution.
However, although the internal structures are different
as compared with the NL3 parametrization,
the densities at which the pasta structure appears and ends are quite
similar.

In tables  \ref{tabNL3} and \ref{tabTW} we have also included the results for
$\beta$-equilibrium matter with trapped neutrinos and a lepton fraction $Y_l=0.4$ at $T=5$
MeV. It is seen that these results are very similar to the ones obtained 
for $Y_p=0.3$. This is
understandable because, for the pasta phase densities, the proton fraction changes from $\sim$  0.29
for the lowest densities to $\sim 0.32$ for the largest ones. A  proton fraction smaller than
0.3 at the
onset of the pasta phase explains the onset of the pasta at  slightly larger densities. On the
other hand a proton fraction $Y_p>0.3$ at the upper limit of the pasta phase corresponds in the
TF calculation to a slightly larger transition density than the $Y_p=0.3$ value, 
and lies between the value obtained for $Y_p=0.5$ and $Y_p=0.3$. 
In the CP calculation the result also lies between  $Y_p=0.5$
and $Y_p=0.3$ but is instead smaller 
 than the $Y_p=0.3$ value because CP does not allow  for the rearrangement of 
the proton distribution.

The comparison between Tables \ref{tabNL3} and \ref{tabTW} is more easily done analysing Fig. \ref{fig2}. 
In Fig. \ref{fig2} we compare the  density range for which each pasta 
configuration exists within NL3 and TW. The thick lines stand for the Thomas-Fermi calculation,
the thin ones for the CP method. Full lines represent TW and dahsed ones NL3.     

For symmetric matter the main difference is the the appearance of the different 
phases at slightly smaller densities within NL3. 
The largest differences occur for $Y_p=0.3$: NL3 has no slab phase at T=7 and 8 MeV 
and quite large tube and bubble phases.

It has been discussed in \cite{oyamatsu07,watanabe08} that the characteristics 
of the pasta phase are strongly related with the density dependence of the 
symmetry energy. In particular, in \cite{watanabe08}  the pasta phases have 
been calculated using quantum molecular
dynamics. The authors have  considered two models, and have obtained  a quite different
structure for the pasta phase of both models. The slab phase is very small in one of the models
and although larger in the other, it is one of the phases that first disappears when the
temperature is raised. 

A large value of the slope of the symmetry energy $L$ means a smaller 
symmetry energy at sub-saturation densities. Therefore,
since TW has a much smaller symmetry energy slope,  $L=55$ MeV for TW and  
$L=118$ MeV for NL3,
at sub-saturation densities its symmetry energy is larger and the neutron gas equation of state
has a larger energy. As a consequence, neutrons do not drip so easily, the surface tension is
larger and the surface is less diffuse. In Table \ref{tab1} we have included the value of the
surface tension $\sigma$ of TW and NL3 for the proton fractions 0.5 and 0.3.  It is seen that TW has a
much larger $\sigma$ for $Y_p=0.3$ than NL3  which is in accordance with its symmetry energy
slope. We may now understand the different properties of the TW pasta: the larger surface
energy explains the fact that all the pasta phases extend to larger densities. 
For non-symmetric matter as for $Y_p=0.3$ this effect is even stronger. This explains some of the
largest differences between NL3 and TW as pointed out before, namely the fact 
that   NL3 has no slab phase above T=7 MeV.
As  already noticed in \cite{pasta1,maru2010}  the extension of the pasta phase decreases with
the increase of the temperature.
This is expected because the surface tension decreases with temperature.

\begin{figure*}[htb]
\begin{center}
\begin{tabular}{cc}
\includegraphics[width=0.45\linewidth]{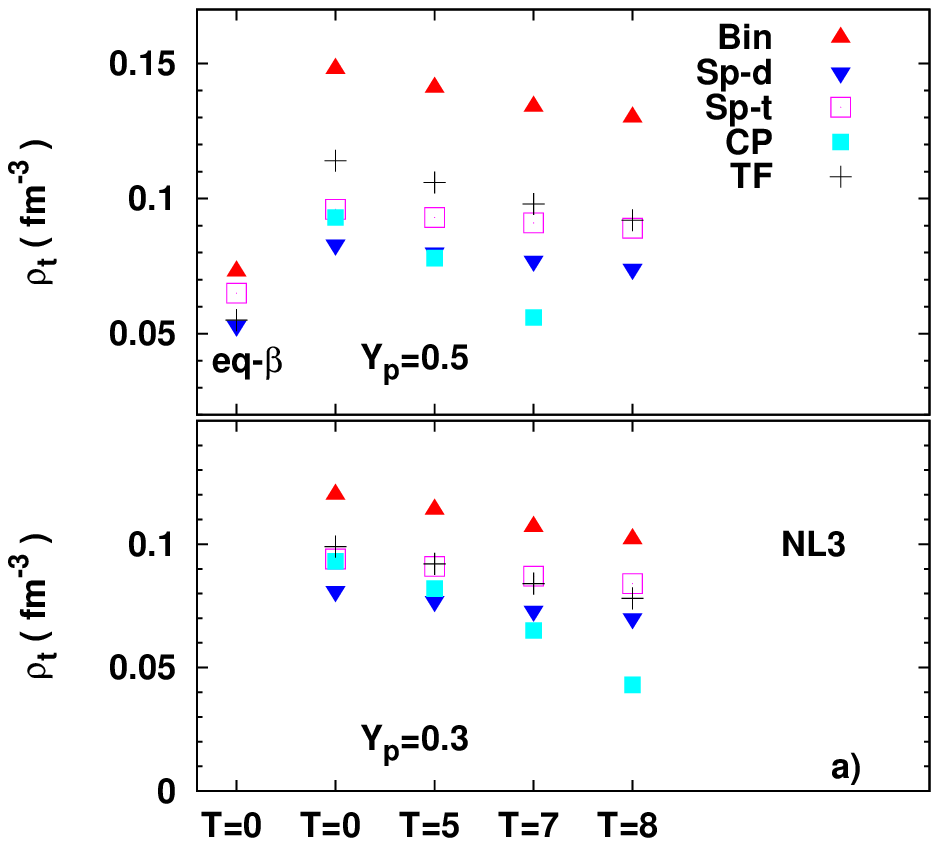} &
\includegraphics[width=0.45\linewidth]{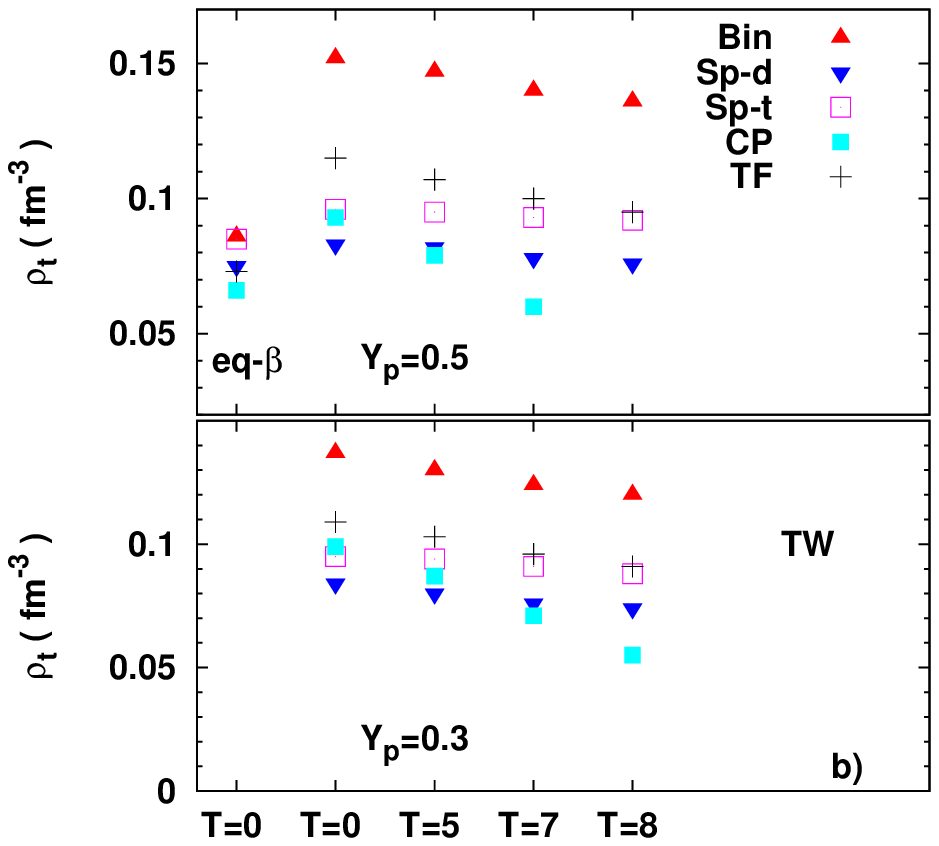} \\
\end{tabular}
\end{center}
\caption{Transition density, for several temperatures $T=0,$ 5, 7 and 8 MeV 
and proton fractions $Y_p=0.5$ (upper plots), $0.3$ (lower plots) 
and $\beta$-equilibrium (eq-$\beta$) at T=0 (left part of upper plots),  
obtained using different methods
within the models a) NL3 and b) TW. 
The methods used are: estimation from binodal (Bin), estimation from 
dynamical spinodal (Sp-d) estimation from thermodynamical spinodal (Sp-th), 
calculation using the coexisting-phases method (CP) and calculation using a 
Thomas-Fermi method (TF).
}
\label{fig1}
\end{figure*}

Fig. \ref{fig2} also allows for a clear comparison between the CP and TF methods: the phases
droplets, rods and slabs are  stable until larger densities with  CP. On the other hand, the
phases tubes and bubbles do not appear as the ground-state configuration since they have
a higher free energy than the homogeneous matter.
Other important differences are the smaller extension of the pasta phase 
within CP (it starts at larger densities and finishes at smaller densities), 
and the disappearance of the pasta phase at smaller temperatures. A smaller extension of the pasta phase within the CP method 
is partially  due to  non-selfconsistent treatment of the Coulomb interaction  \cite{maruyama}, which prevents the rearrangement of the proton distributions (see also \cite{pasta_alpha}). This is more strongly seen for the $Y_p=0.5$ when Debye screening effects are stronger.  We have calculated the critical temperatures within the  CP calculation, above which the pasta phase does not exist.  For NL3 we have 
$T_c^{CP}=7.73,\, (8.09)$ MeV for $Y_p=0.5, \, (0.3)$, for TW $T_c^{CP}=7.95,\, (8.42)$ MeV for $Y_p=0.5, \, (0.3)$. The CP critical temperature occurs when the free energy of the homogeneous phase is smaller than the free energy of the pasta phase, and is not defined by a zero surface tension, as in TF. At and above the critical temperature,  pasta clusters still exist within CP calculation  but with a larger free energy than homogeneous matter.

Finally, it is interesting to compare the prediction of different methods for 
the crust-core (non-homogeneous--homogeneous) transition density. 
The thermodynamical binodal and spinodal of proton-neutron $(pn)$ matter give a good estimation of the extension of the pasta
phase in nuclear matter. The binodal surface is defined  in the $\rho,\, Y_p,\, T$ space
as the surface where the gas and liquid phases coexist and it defines an  upper limit for the
extension of the pasta phase since it does not take into account neither Coulomb nor finite
size effects.  The thermodynamical spinodal defines the surface in the $\rho,\, Y_p,\, T$
space where the curvature of the free-energy of $pn$ matter becomes negative, 
and gives a rough
estimation of the lower limit of the pasta phase extension. Also in this case
nor electron effects neither finite range effects are taken into 
account. 
The transition density is obtained from the intersection between the
$\beta$-equilibrium EoS and the spinodal or binodal surfaces.
It is possible to get a better estimation of the lower
limit of the pasta phase extension if, instead of the thermodynamical spinodal, the dynamical
spinodal is calculated. This surface is defined by the density, proton fraction and temperature
for which  matter becomes unstable to  small density fluctuations and is calculated including
electrons, the Coulomb interaction and finite range effects. 
 It has been shown in \cite{oyamatsu07,pasta1,pasta2,xu09}
 that the thermodynamical method gives a good estimation  of the transition density for cold
$\beta$-equilibrium matter, although a bit too large.
For a fixed temperature, the binodal, thermodynamical and dynamical spinodals touch each other (or almost touch) at the
critical point \cite{muller95}: this corresponds to a quite small proton fraction
and is close to the crust-core transition density. However, for less asymmetric
matter as the one that occurs in stellar matter with trapped neutrinos, 
 the binodal occurs at quite larger densities than the spinodals. This is clearly
seen in Fig. \ref{fig1}
comparing  the beta-equilibrium results for cold stellar matter with the results for
$Y_p=0.5$ or 0.3.

For densities above the binodal 
surface, matter is homogeneous. For densities below the dynamical spinodal 
matter is non-homogeneous. Between this two surfaces we may find matter in
a metastable configuration. The most probable configuration is the one with the smallest free
energy density. The results given in Tables \ref{tabNL3}  and \ref{tabTW} 
refer to the
energetically favoured configurations. However, we may argue that at finite 
temperature there
may be a coexistence of several configurations with larger or smaller probability according to
the corresponding free energy density. This partially justifies the large differences between
the TF and CP results: although the ground-state configurations are different, there is not a
large energy difference between the configurations and, taking into account an average over all
the possible configurations with a correct probability factor would give similar predictions.  

In Fig. \ref{fig1} we compare the transition density from a non-homogeneous phase to a homogeneous 
phase obtained from the binodal surface \cite{muller95}, the dynamical spinodal
surface \cite{pethick95,douchin00,providencia06,ducoin08,ducoin08a,xu09}, the
thermodynamic spinodal surface \cite{baoli02,avancini06,xu09} and from the two methods we have used for the calculation of the
pasta phase. As expected the TF result lies always between the results obtained from  the
dynamical spinodal and the binodal surfaces. This is a self-consistent method that should
satisfy these two constraints. This is no longer true for the CP calculation, 
which, for larger temperatures, fails with predictions that lie below the dynamical spinodal result. It is
impressive that the thermodynamical-spinodal result gives so good results when it does not take
into account neither the surface nor the Coulomb effects. These are good news because it allows a
quite safe prediction from a simple calculation.
Finally, it should also be pointed out that for $\beta$-equilibrium matter all methods except CP give similar results. This is due to the occurrence of the transition density close the critical point where the spinodal and binodal surfaces touch and the pressure on these surfaces is maximum.

\begin{table*}[thb]
\caption{Pasta phases within NL3 for several temperatures and isospin asymmetries.
 The onset densities of the various phases (droplets, rods, slabs, tubes and bubbles) are shown together with
the pressure of the homogeneous phase at the lower   ($P^1$) and upper  ($P^2$)
border of the pasta phase. The results have been obtained within the coexisting phases method (CP) and
the Thomas Fermi approximation (TF).}
\begin{center}
\begin{tabular}{ll cccccc c}
\hline
&&\multicolumn{6}{c}{onset density (fm$^{-3}$)} &  $P^1~/~P^2$\\
& EOS  & droplet  &   rod    &  slab   &  tube     & bubble  & hom& (MeV~fm$^{-3}$) \\
\hline
%
%  T=0
%
$T=0$ MeV\\
$Y_p=0.5$\\
   & TF  & 0.000& 0.025    & 0.043  & 0.073     &  0.089   & 0.114 &   - / 2.75     \\
   & CP  & 0.000& 0.032    &  0.053 &   -       &     -    & 0.093 & - / 1.88  \\
$Y_p=0.3$\\
   &TF & 0.000&  0.019   & 0.036 & 0.059   &  0.071   &0.099  &  - / 1.11      \\
   &CP   & 0.000 & 0.030    &  0.048 & 0.084     &     -    & 0.093 &- / 0.96 \\
$\beta$-equilibrium\\
   & TF  & 0.000 & -    & - &  -    & -    &  0.055  &  - /0.26   \\
   &CP   & - &  -   & -  &  -   &     -    & - &  -/- \\
\hline
\hline
%
%  T=5
%
$T=5$ MeV\\
$Y_p=0.5$\\
   & TF  &  0.0020     & 0.026    &  0.043 &  0.067    &  0.081   & 0.106 &   ~  0.024$~/~$2.55     \\
   & CP  &  0.0035     & 0.031    &  0.051 &   -       &     -    & 0.078 &  ~0.046$~/~$1.53  \\
$Y_p=0.3$\\
   &TF &  0.0022     & 0.024    &  0.038 & 0.053     &  0.066   & 0.092 &    ~  0.018$~/~$1.08     \\
   &CP   &  0.0029     & 0.032    &  0.049 & 0.081     &     -    & 0.082 & ~ 0.025$~/~$0.85\\
$Y_l=0.4$\\
 & TF&0.0024  & 0.023 & 0.036 & 0.054 & 0.065 & 0.093  & ~0.023$~/~$5.24 \\
 & CP&0.0033  & 0.031 & 0.049 &  -    & -  & 0.080&  0.024 / 1.43   \\
\hline
%
%     T=7
%
$T=7$ MeV\\
$Y_p=0.5$\\
   & TF  &  0.0048     &  0.026   &  0.042 &    0.054  & 0.072    & 0.098 &    ~ 0.076$~/~$2.37     \\
   & CP  &  0.013      &  0.030   & 0.049  &   -       &     -    & 0.056 & ~ 0.22$~/~$1.06  \\
$Y_p=0.3$\\
   & TF  &  0.0052     &  0.025   &   -    &   0.036  &  0.059   & 0.084 &   ~ 0.056$~/~$1.02     \\
   &CP   & 0.0107      & 0.032    & 0.048  &   -       &     -    & 0.065 & ~0.112$~/~$0.67 \\
\hline
%
%    T=8
%
$T=8$ MeV\\
$Y_p=0.5$\\
   & TF  & 0.0069      &  0.025   &   -    &  0.036   &    0.063  &  0.092&    ~ 0.121$~/~$2.23     \\
   & CP  &    -        &    -     &    -   &   -      &     -    &    -  &    ~ -$~/~$-     \\
$Y_p=0.3$\\
   & TF  &  0.0076    &   0.026  &  - &  0.033    &  0.055   &   0.078    &     ~ 0.090$~/~$0.96      \\
   &CP   &  0.0239     & 0.032    &   -    &   -       &     -    &  0.043& ~0.253$~/~$0.44 \\

\hline
\end{tabular}
\end{center}
\label{tabNL3}
\end{table*}

\begin{table*}[h]
\caption{Pasta phases within TW for several temperatures and isospin asymmetries.
 The onset densities of the various phases (droplets, rods, slabs, tubes and bubbles) are shown together with
the pressure of the homogeneous phase at the lower   ($P^1$) and upper  ($P^2$)
border of the pasta phase. The results have been obtained within the coexisting phases method (CP) and
the Thomas Fermi approximation (TF).}
\begin{center}
\begin{tabular}{ll cccccc c}
\hline
&&\multicolumn{6}{c}{onset density (fm$^{-3}$)} &  $P^1~/~P^2$\\
& EOS  & droplet  &   rod    &  slab   &  tube     & bubble  & hom& (MeV~fm$^{-3}$) \\
\hline
%
%  T=0
%
$T=0$ MeV\\
$Y_p=0.5$\\
   & TF  & 0.000 &0.025     & 0.043  &  0.075    & 0.090    & 0.115 &  - /2.77     \\
   & CP  &0.000 & 0.033    &  0.054 &   -       &     -    & 0.093 & - /1.85  \\
$Y_p=0.3$\\
   &TF & 0.000&   0.021  & 0.040 &  0.071  &  0.085   & 0.109 &    -/   1.22  \\
   &CP   & 0.000 & 0.031    &  0.051 & 0.89    &     -    & 0.099 &  -/ 0.96\\
$\beta$-equilibrium\\
   & TF  & 0.000 & -    & - &  -    & -    &  0.073  &  - / 0.36  \\
   &CP   & 0.003 &  -   & -  &  -   &     -    & 0.066 &  0.007/ 0.29 \\
\hline
$T=5$ MeV\\
$Y_p=0.5$\\
   & TF  &    0.0022   &   0.026   &  0.043 &   0.071   &  0.084   & 0.107 &  ~ 0.027$~/~$2.53   \\
   & CP  &   0.0038    &  0.032   &  0.052 &   -       &     -    & 0.079 &  ~0.050$~/~$1.55  \\
$Y_p=0.3$\\
   &TF   &   0.0024    & 0.025    & 0.042 & 0.068    &  0.080  & 0.103 &    ~0.020$~/~$1.20   \\
   &CP   &  0.0032     & 0.033    &  0.052 &   -        &     -    & 0.087 & ~0.028$~/~$0.86\\
$Y_l=0.4$\\
 & TF&0.0026  & 0.025 & 0.042 & 0.068 & 0.081 & 0.104 & ~0.026$~/~$1.64 \\
 & CP&0.0035  & 0.033 & 0.051 &   -    &  -   &  0.085  & ~0.036$~ /~$1.13 \\
\hline
$T=7$ MeV\\
$Y_p=0.5$\\
   & TF  &  0.0051     &  0.027   & 0.043  &  0.063    & 0.077    & 0.100 &   ~0.082$~/~$2.41    \\
   & CP  &  0.0130     & 0.031    & 0.051  &   -       &     -    & 0.060 &  ~0.225$~/~$1.16  \\
$Y_p=0.3$\\
   & TF   &  0.0055    &   0.027  & 0.042 &  0.062     & 0.074    & 0.096 &  ~0.061$~/~$1.16    \\
   &CP   &  0.0110     & 0.033    & 0.051  &   -       &     -    & 0.071 &  ~0.120$~/~$0.72\\
\hline
$T=8$ MeV\\
$Y_p=0.5$\\
   & TF &  0.0073      &  0.027   & 0.043  &  0.056    & 0.072    &  0.095 &   ~0.130$~/~$2.32   \\
   & CP  &    -        &    -     &    -   &   -       &     -    &    -   &   ~  -$~/~$-     \\
$Y_p=0.3$\\
   & TF  &  0.0079    &   0.030   & 0.043  &  0.057    & 0.069    &  0.091 &    ~ 0.096$~/~$1.13   \\
   &CP   &  0.0207     &  0.033   &   0.050&   -       &     -    & 0.055 &   ~0.234$~/~$0.57 \\
\hline
\end{tabular}
\end{center}
\label{tabTW}
\end{table*}

\section{Conclusions}\label{sec:conclusions}

In the present work we have calculated the pasta phase at zero and finite
temperature applying two different methods already used in previous works: 
the more naive coexisting-phases method based on the Gibbs 
construction
and the self-consistent Thomas Fermi method. For the first method we need to 
know the surface tension of the models as a function of temperature and proton fraction. This quantity was
parametrized from a Thomas Fermi calculation for semi-infinite nuclear matter \cite{pasta_alpha}. We
had already compared the two models at zero temperature \cite{pasta1,pasta2} but for the CP
calculation we had used a parametrization for the surface tension  obtained from Skyrme
forces. Since the appearance of the pasta phases has a strong dependence on 
this quantity the comparison between the models at zero temperature was not 
conclusive.

We have considered two different relativistic nuclear models: NL3, a parametrization of the NLWM,
and TW, a parametrization of the density dependent hadronic model. These two models have quite
different behaviors both in the isoscalar and isovector channels. 
In particular,  TW has a quite small slope of the symmetry energy at saturation, $L=$ 55
MeV, while this value is 118 MeV for NL3. The properties of the models are clearly reflected in
the pasta phase structure as discussed in \cite{watanabe08}.

We conclude that while the CP method allows the determination of the 
overall trends of both models, it fails at a more detailed and quantitative
level. The main trends observed are: the model having a larger
surface tension predicts larger density ranges for the pasta phases; 
the pasta phase extension decreases with temperature; some of the 
less stable  pasta structures may disappear for larger temperatures and/or 
isospin asymmetries. We have also shown
that the overall conclusions obtained within the Thomas Fermi calculation at finite temperature
agree with the conclusions obtained from a quantum molecular dynamics calculation
\cite{watanabe08}. In fact, it was shown that the structure of the pasta is sensitive to the
model, namely the density dependence of the symmetry energy. A model with a large symmetry
energy slope such as NL3, has a quite small symmetry energy at sub-saturation densities and this
favours neutron drip. As a consequence the background neutron gas is larger for NL3 at a given
density and   therefore, the  favoured
pasta structures change shape at smaller densities.

We have also compared the crust-core transition density, from a 
non-homogeneous phase to a homogeneous phase, obtained  from the above two 
methods as well as using other methods  often
referred to in the literature. It was also shown that the dynamical
spinodal defines a lower limit while the binodal a higher limit and the 
TF result lies between the two limits. The CP 
methods fails these
constraints both for large temperatures, above T=5 MeV,  and very
asymmetric matter. 
We have obtained the interesting result that the estimates obtained from a 
thermodynamical calculation are very close to the prediction of the
TF calculation.

\section*{ACKNOWLEDGMENTS}
This work was partially supported by Capes/FCT n. 232/09 bilateral
collaboration, by CNPq (Brazil), by FCT and FEDER (Portugal) under the projects
PTDC/FIS/64707/2006, CERN/FP/109316/2009 and SFRH/BPD/64405/2009 
and  by Compstar, an ESF Research Networking
Programme.

\section*{Appendix}

In this appendix we calculate Eq. (\ref{sigcp}) for the surface tension $\sigma$,
following very closely reference \cite{dmcp}. 
The system is composed by $np$ matter and the density depends only
upon the $z$ coordinate.
Notice that no Coulomb field is included in the calculation of $\sigma$.
We start from the grand-canonical potential, Eq. (\ref{grand1}), where
\begin{equation}
\Omega=  \int\d V \left\{\frac{1}{2}\left [
(\nabla \phi_0)^2 -(\nabla V_0)^2 - (\nabla b_0)^2\right] - V_{ef}\right\}
\label{grand3}
\end{equation}
with
\begin{eqnarray}
V_{ef}&=&
-\frac{1}{2} \left[ 
m_s^2 \phi_0^2 + \frac{2}{3!} \kappa \phi_0^3 + \frac{2}{4!} \lambda \phi_0^4\right.\nonumber\\
&&
\phantom{mmm}\left.-m_v^2 V_0^2-m_\rho^2 b_0^2 + 2 \Sigma_0^R \rho \right]\nonumber\\
&+& 2 T \sum_{i=p,n} \int \frac{\d^3p}{(2\pi)^3} \left[  
\ln (1 + e^{-(\epsilon^* - \nu_i)/T})\right.\nonumber\\
&+&\left. \ln (1 + e^{-(\epsilon^* + \nu_i)/T}) \right] \label{vef}.
\end{eqnarray}
The equations of motion for the the meson fields are obtained by minimizing $\Omega$ with
respect to each field, 
\begin{eqnarray}
\nabla^2 \phi_0 &=& \frac{\d^2 \phi_0}{\d z^2} 
=-\frac{\partial V_{ef}}{\partial\phi_0},
\label{phi2}\\
\nabla^2 V_0 &=&\frac{\d^2 V_0}{\d z^2} 
= \frac{\partial V_{ef}}{\partial V_0},
\label{V02}\\
\nabla^2 b_0 &=&\frac{\d^2 b_0}{\d z^2} 
=\frac{\partial V_{ef}}{\partial b_0}.
\label{b02}
\end{eqnarray}
Using the relation
$$\frac{\d^2\, W}{\d z^2}=\frac{d}{dz}\left(\frac{d\, W}{dz}\right) =
\frac{1}{2} \frac{d}{dW}\left(\frac{d\,
W}{dz}\right)^2,$$
where $W=\phi_0,V_0,b_0$,
we obtain
\begin{eqnarray}
\frac{\partial }{\partial\phi_0} \left[\frac{1}{2}
\left(\frac{d\, \phi_0}{dz}\right)^2 +V_{ef}\right]&=&0\nonumber\\
\frac{\partial }{\partial V_0} 
\left[\frac{1}{2}\left(\frac{d\, V_0}{dz}\right)^2 -V_{ef}\right]&=&0\nonumber\\
\frac{\partial }{\partial b_0} 
\left[\frac{1}{2}\left(\frac{d\, b_0}{dz}\right)^2 -V_{ef}\right]&=&0~.
\end{eqnarray}
Summing adequately the three equations (the second and third equations multiplied by -1), we
get
\begin{equation}
\delta\left[\frac{1}{2}\left(\frac{d\, \phi_0}{dz}\right)^2
-\left(\frac{d\, V_0}{dz}\right)^2
 -\left(\frac{d\, b_0}{dz}\right)^2 
+V_{ef}\right]=0.
\end{equation}
This is equivalent to saying that 
\begin{equation}
\frac{1}{2}\left[\left(\frac{d\, \phi_0}{dz}\right)^2
-\left(\frac{d\, V_0}{dz}\right)^2
 -\left(\frac{d\, b_0}{dz}\right)^2 \right]
+V_{ef}={\cal{C}},\label{c}
\end{equation}

where ${\cal{C}}$ is a constant which corresponds to the bulk contribution to the grand canonical
potential density and can be identified with the pressure $P$.
Replacing Eq. (\ref{c}) into (\ref{grand3}) we obtain
\begin{equation}
\Omega=  \int\d V \left [\left(\frac{d\, \phi_0}{dz}\right)^2
-\left(\frac{d\, V_0}{dz}\right)^2
 -\left(\frac{d\, b_0}{dz}\right)^2
\right]- {\cal{C}}V.
\label{grand4}
\end{equation}

The surface energy is obtained from the free energy of a system with a fixed number of
particles $N=N_p+N_n$, in which a cluster of arbitrary size exists
in the background of the vapor phase. 
The free energy reads
\begin{eqnarray}
F&=&\Omega
 +\mu_p N_p +\mu_n N_n,\nonumber\\
&=&S\int_{-\infty}^{\infty} \d z
\left[ \left(\frac{\d \phi_0}{\d z}\right)^2-
\left(\frac{\d V_0}{\d z}\right)^2 -
\left(\frac{\d b_0}{\d z}\right)^2
\right]\nonumber\\
&&- {\cal{C}}V +\mu_p N_p +\mu_n N_n.
\end{eqnarray}                      
For a cluster of 
volume $V$ and surface $S$, we have
\begin{equation}
F=S \sigma - {\cal{C}}V + \mu_p N_p +\mu_n N_n. \label{free}
\end{equation}
The surface energy per unit area of this cluster
is
\begin{equation}
\sigma=\int_{-\infty}^\infty \d z \left[ \left(\frac{\d \phi_0}{\d z}\right)^2- 
\left(\frac{\d V_0}{\d z}\right)^2 -
\left(\frac{\d b_0}{\d z}\right)^2
\right].
\label{sig}\end{equation}
Eq. (\ref{free}) can be rewritten in the form of Eq. (3.14) of reference \cite{centel}
\begin{eqnarray}
 \sigma = \int_{-\infty}^\infty dz \left[ {\cal F}(z) -{\cal F}_{g}  - \mu_p (\rho_{p}(z)-\rho_{p,g})
\right.\nonumber\\
\phantom{mmmmmm}\left. -\mu_n  (\rho_{n}(z)-\rho_{n,g})\right]. \label{free1}
\end{eqnarray}
where
${\cal F}_g=-P+ \mu_p \rho_{p,g} + \mu_n \rho_{n,g}$, $\rho_{p,g}$ and $\rho_{n,g}$
are the free energy density, the proton density and the neutron density
of the gas. We have checked numerically the equivalence
between eqs. (\ref{sig}) and (\ref{free1}).

\include{err}

\end{document}

%% file: err.tex
\section*{Erratum: Warm pasta phase in the Thomas-Fermi approximation [Phys.\ Rev.\ C 82, 055807 (2010)]}
\pagenumbering{roman}
\setcounter{figure}{0}
\setcounter{table}{0}

\author{Sidney S. Avancini}
\affiliation{Depto de F\'{\i}sica - CFM - Universidade Federal de Santa
Catarina  Florian\'opolis - SC - CP. 476 - CEP 88.040 - 900 - Brazil}
\author{Silvia Chiacchiera}
\affiliation{Centro de F\'{\i}sica Computacional - Department of Physics -
University of Coimbra, P3004 - 516, Coimbra, Portugal}
\author{D\'ebora P. Menezes}
\affiliation{Depto de F\'{\i}sica - CFM - Universidade Federal de Santa
Catarina  Florian\'opolis - SC - CP. 476 - CEP 88.040 - 900 - Brazil}
\author{Constan\c ca Provid\^encia}
\affiliation{Centro de F\'{\i}sica Computacional - Department of Physics -
University of Coimbra, P3004 - 516, Coimbra, Portugal}

\maketitle

\vspace{0.50cm}
PACS number(s): {21.65.-f, 24.10.Jv, 26.60.-c, 95.30.Tg}
\vspace{0.50cm}

There was an error in the numerical code for the CP method that resulted in 
using the zero temperature surface tension for all the cases, instead of 
the temperature dependent one. 
Also, applying the CP method, we found that in the
surface energy parametrization given in Ref. \cite{pasta_alpha2} 
the global proton fraction should be used.
As a result, the agreement between the more ``na\"{\i}ve'' CP and
the TF calculations has improved. The figures and the table including the correct 
values for CP are shown below.

\begin{figure*}[htb]
\begin{center}
\begin{tabular}{cc}
\includegraphics[width=0.45\linewidth]{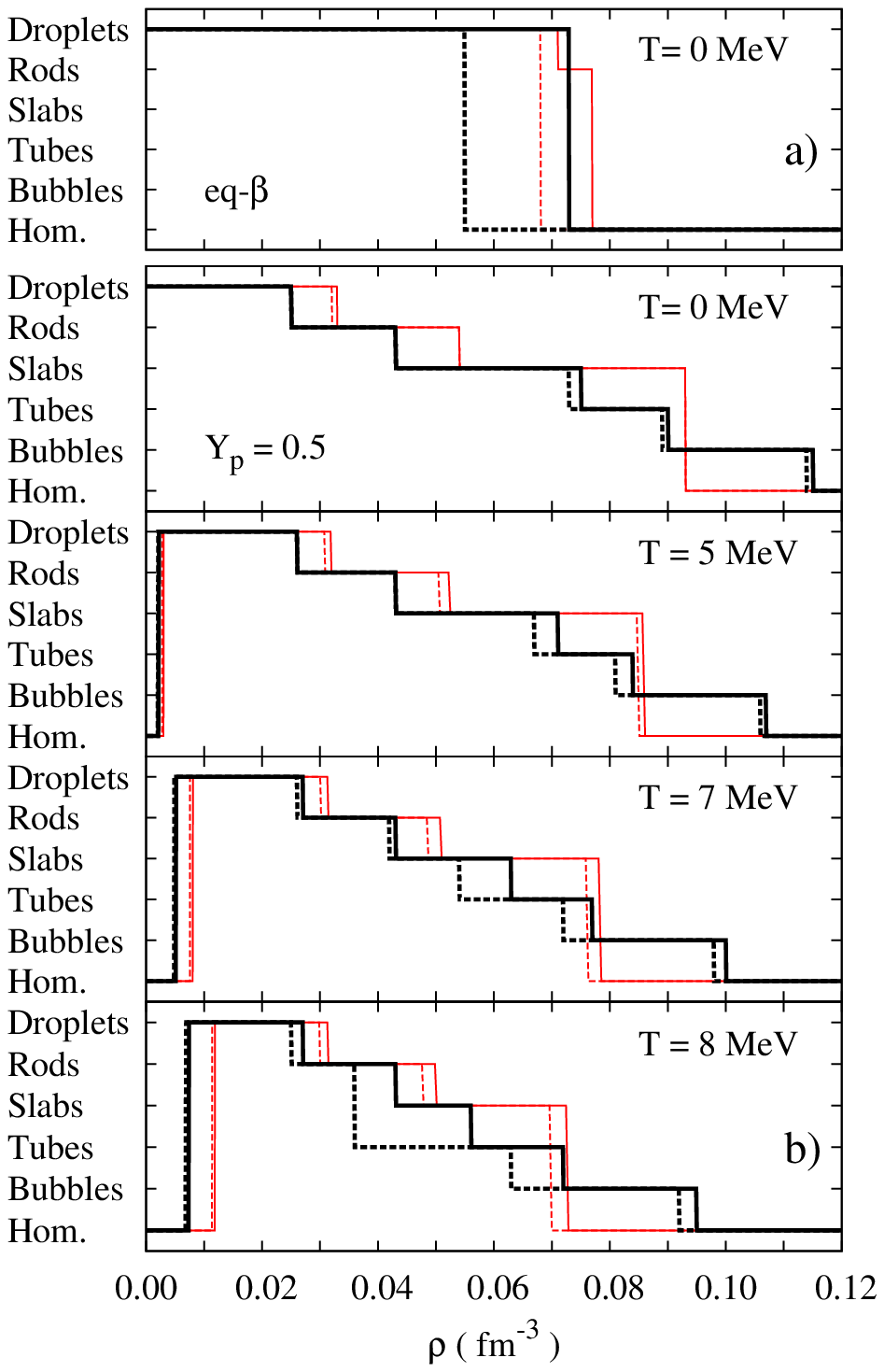} &
\includegraphics[width=0.45\linewidth]{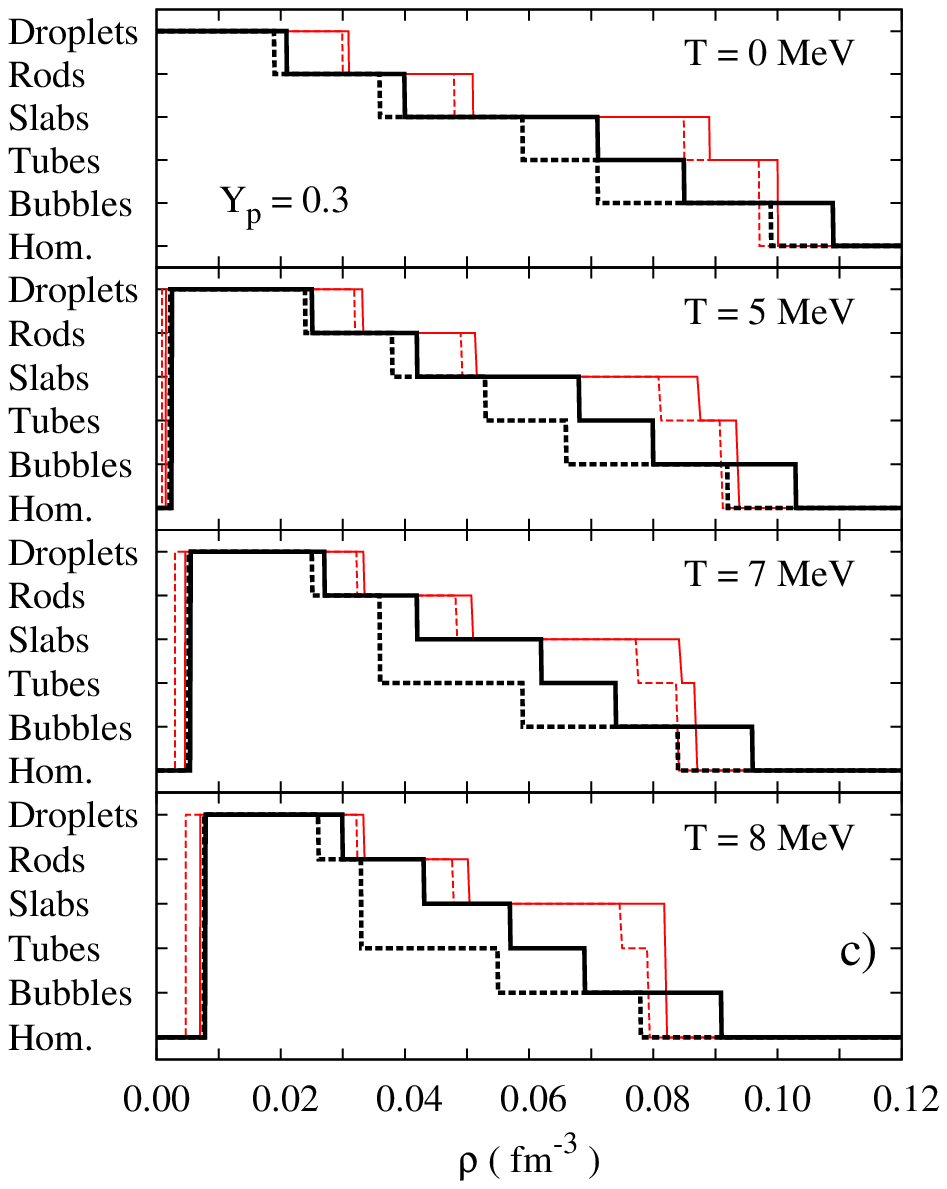} \\
\end{tabular}
\end{center}
\caption{Pasta phases: comparison between  NL3 (dashed line)  and TW (full line)
and the methods CP and TF for a) $\beta$-equilibrium cold stellar  matter; 
b) $Y_p=0.5$, c) $Y_p=0.3$. The thick lines refer to the Thomas-Fermi (TF) calculation and the thin red
lines to the coexisting-phases (CP) calculation.}
\label{fig1}
\end{figure*}

\begin{figure*}[htb]
\begin{center}
\begin{tabular}{cc}
\includegraphics[width=0.45\linewidth]{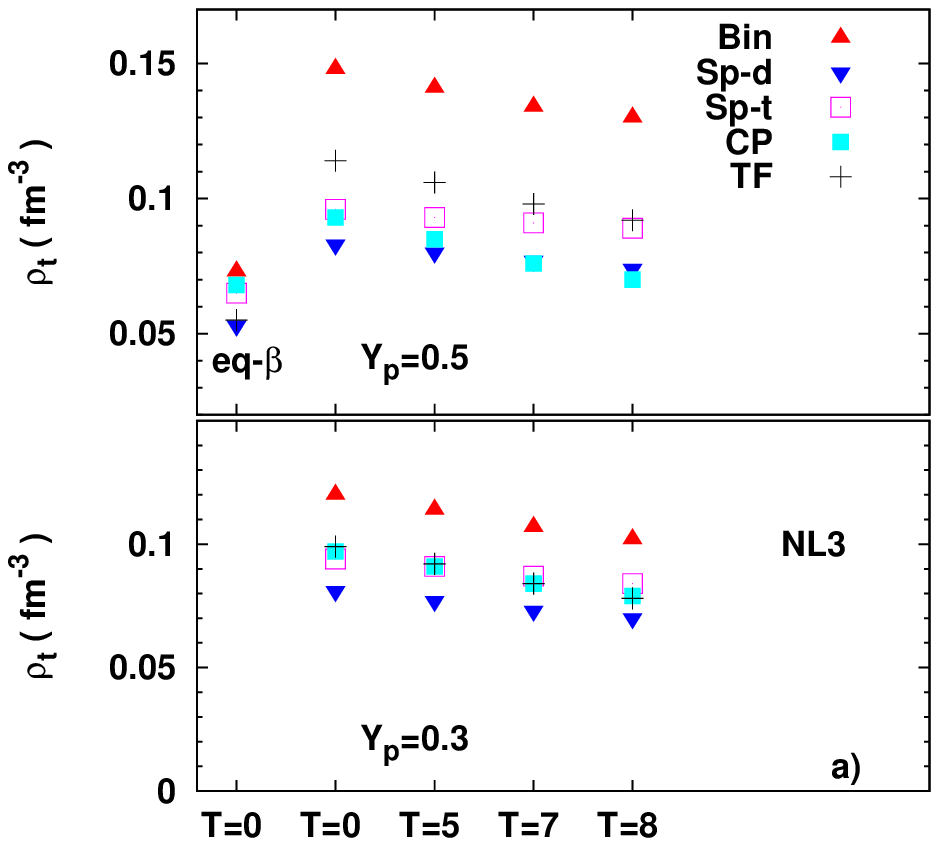} &
\includegraphics[width=0.45\linewidth]{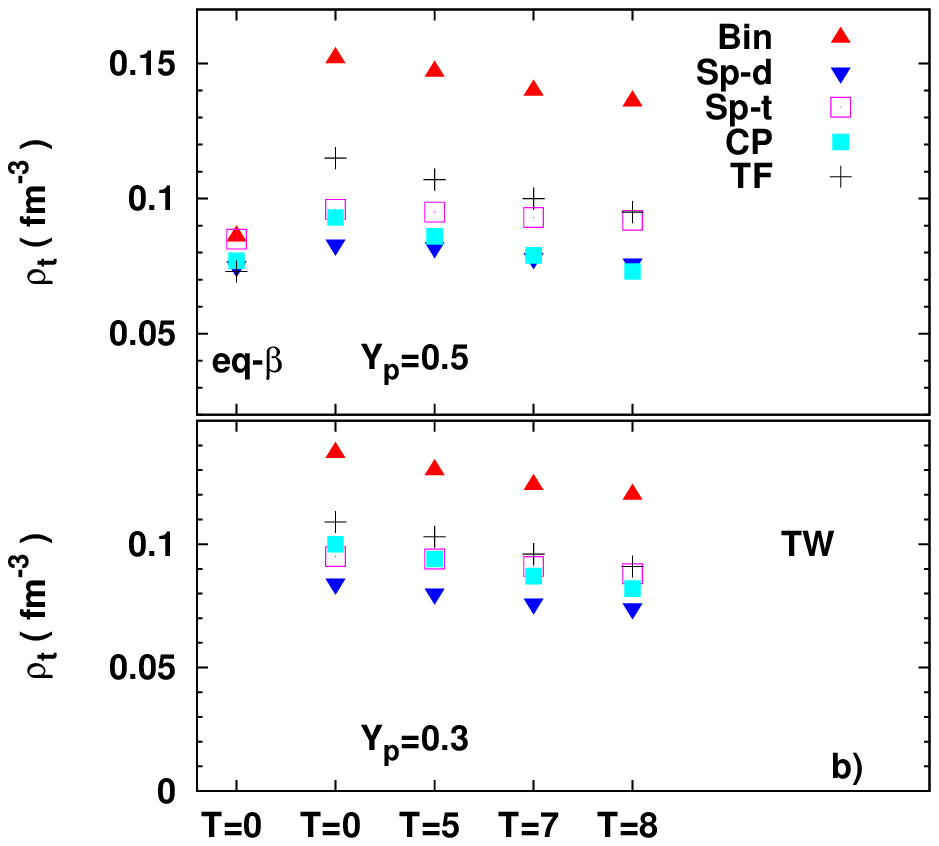} \\
\end{tabular}
\end{center}
\caption{Transition density, for several temperatures $T=0,$ 5, 7 and 8 MeV 
and proton fractions $Y_p=0.5$ (upper plots), $0.3$ (lower plots) 
and $\beta$-equilibrium (eq-$\beta$) at T=0 (left part of upper plots),  
obtained using different methods
within the models a) NL3 and b) TW. 
The methods used are: estimation from binodal (Bin), estimation from 
dynamical spinodal (Sp-d) estimation from thermodynamical spinodal (Sp-th), 
calculation using the coexisting-phases method (CP) and calculation using a 
Thomas-Fermi method (TF).
}
\label{fig2}
\end{figure*}

In the following, some specific comments we made are corrected.
In the CP approximation, the ``pasta'' phase is found in all the cases considered:
the critical temperature for its existence is therefore higher than $8$ MeV.
The region in which the ``pasta'' phase exists is still generally smaller in the CP 
approach than in the TF one, but some exceptions are found (see Fig.~\ref{fig1}).
Concerning the transition density (see Fig.~\ref{fig2}), the CP results are now 
much closer to the TF ones. Moreover, for temperatures lower than 7
MeV, they satisfy the constraints imposed by the 
two spinodals.

\begin{table*}
\caption{
Pasta phases within NL3 and TW for several temperatures and isospin asymmetries.
 The onset densities of the various phases (droplets, rods, slabs, tubes and bubbles) are shown together with
the pressure of the homogeneous phase at the lower   ($P^1$) and upper  ($P^2$)
border of the pasta phase. 
The results have been obtained within the coexisting-phases method (CP), corrected according to the text explanation, 
and the Thomas-Fermi approximation (TF), coincident with results in Phys.\ Rev.\ C 82, 055807 (except for two misprints 
for $P^2$ at  $T=5$ MeV and with $Y_l=0.4$). }
\begin{center}
\begin{tabular}{cc ccccccc c ccccccc}
\hline\hline
          & EOS & \multicolumn{7}{c}{NL3} & & \multicolumn{7}{c}{TW}\\
\cline{3-9}\cline{11-17}
          &     & \multicolumn{6}{c}{Onset density (fm$^{-3}$)}       &  $P^1~/~P^2$  &&\multicolumn{6}{c}{Onset density (fm$^{-3}$)}          &  $P^1~/~P^2$\\
\cline{3-8}\cline{11-16}
          &     & droplet &  rod   &  slab  &  tube  & bubble & hom   &(MeV/fm$^{3}$) && droplet &  rod  &  slab &  tube  & bubble & hom   & (MeV/fm$^{3}$)\\
\hline
%T=0 MeV
$T=0$ MeV   & & & & & & & & & & & & & & & \\
$Y_p=0.5$ & TF  & 0.000   & 0.025  & 0.043  & 0.073  & 0.089  & 0.114 &   - / 2.75    && 0.000   & 0.025 & 0.043 &  0.075 & 0.090  & 0.115 &    - / 2.77  \\
$Y_p=0.5$ & CP  & 0.000   & 0.032  & 0.054  &   -    &     -  & 0.093 &   - / 1.88    && 0.000   & 0.033 & 0.054 &   -    &  -     & 0.093 &    - / 1.85  \\ 
$Y_p=0.3$ & TF  & 0.000   & 0.019  & 0.036  & 0.059  &  0.071 & 0.099 &   - / 1.11    && 0.000   & 0.021 & 0.040 &  0.071 & 0.085  & 0.109 &    - / 1.22  \\
$Y_p=0.3$ & CP  & 0.000   & 0.030  & 0.048  & 0.085  &     -  & 0.097 &   - / 0.96    && 0.000   & 0.031 & 0.051 &  0.089 &   -    & 0.100 &    - / 0.96  \\ 
$\beta$-equil.& TF  & 0.000   & -      & -      &  -     & -      & 0.055 &   - / 0.26    && 0.000   & -     & -     &  -     &  -     & 0.073 &    - / 0.36  \\
$\beta$-equil.& CP  & 0.000   &  -     &  -     &  -     &     -  & 0.068 &  -  / 0.49    && 0.000   & 0.071 & -     &  -     &   -    & 0.077 &    - / 0.41  \\ 
%\\
%T=5 MeV
$T=5$ MeV & & & & & & & & & & & & & & & \\
$Y_p=0.5$ & TF  & 0.0020  & 0.026  & 0.043  &  0.067 &  0.081 & 0.106 & 0.024 / 2.55  && 0.0022  & 0.026 & 0.043  & 0.071 &  0.084 & 0.107 &  0.027 / 2.53  \\
$Y_p=0.5$ & CP  & 0.0028  & 0.031  & 0.051  &   -    &     -  & 0.085 & 0.035 / 1.76  && 0.0030  & 0.032 & 0.052  &   -   &     -  & 0.086 &  0.039 / 1.76  \\
$Y_p=0.3$ & TF  & 0.0022  & 0.024  & 0.038  &  0.053 &  0.066 & 0.092 & 0.018 / 1.08  && 0.0024  & 0.025 & 0.042  & 0.068 &  0.080 & 0.103 &  0.020 / 1.20  \\
$Y_p=0.3$ & CP  & 0.0009  & 0.032  & 0.049  &  0.081 &     -  & 0.091 & 0.007 / 1.06  && 0.0015  & 0.033 & 0.052  & 0.088 &   -    & 0.094 &  0.012 / 1.01  \\
$Y_l=0.4$ & TF  & 0.0024  & 0.023  & 0.036  &  0.054 & 0.065  & 0.093 & 0.023 / 1.38  && 0.0026  & 0.025 & 0.042  & 0.068 & 0.081  & 0.104 &  0.026 / 1.64  \\
$Y_l=0.4$ & CP  & 0.0011  & 0.031  & 0.049  &  0.082 & -      & 0.090 & 0.009 / 1.36  && 0.0019  & 0.033 & 0.052  & 0.089 &  -     & 0.092 &  0.018 / 1.30  \\ 
%\\
%T=7 MeV
$T=7$ MeV & & & & & & & & & & & & & & & &\\
$Y_p=0.5$ & TF  & 0.0048  & 0.026  & 0.042  &  0.054 & 0.072  & 0.098 & 0.076 / 2.37  && 0.0051  & 0.027 & 0.043  & 0.063 & 0.077  & 0.100 &  0.082 / 2.41  \\
$Y_p=0.5$ & CP  & 0.0076  & 0.030  & 0.049  &   -    &     -  & 0.076 & 0.124 / 1.62  && 0.0081  & 0.031 & 0.051  &   -   &     -  & 0.079 &  0.135 / 1.65  \\
$Y_p=0.3$ & TF  & 0.0052  & 0.025  &   -    &  0.036 & 0.059  & 0.084 & 0.056 / 1.02  && 0.0055  & 0.027 & 0.042  & 0.062 & 0.074  & 0.096 &  0.061 / 1.16  \\
$Y_p=0.3$ & CP  & 0.0030  & 0.032  & 0.048  &  0.078 &     -  & 0.084 & 0.032 / 1.02  && 0.0046  & 0.033 & 0.051  & 0.085 &     -  & 0.087 &  0.051 / 0.99  \\
%\\
%T=8 MeV
$T=8$ MeV & & & & & & & & & & & & & & & \\
$Y_p=0.5$ & TF  & 0.0069  & 0.025  &   -    &  0.036 & 0.063  & 0.092 & 0.121 / 2.23  && 0.0073  & 0.027 & 0.043  &  0.056 & 0.072 & 0.095 &  0.130 / 2.32  \\
$Y_p=0.5$ & CP  & 0.0114  & 0.030  & 0.048  &   -    &     -  & 0.070 & 0.205 / 1.50  && 0.0119  & 0.031 & 0.050  &   -    &     - & 0.073 &  0.217 / 1.56  \\
$Y_p=0.3$ & TF  & 0.0076  & 0.026  &  -     &  0.033 & 0.055  & 0.078 & 0.090 / 0.96  && 0.0079  & 0.030 & 0.043  &  0.057 & 0.069 & 0.091 &  0.096 / 1.13  \\
$Y_p=0.3$ & CP  & 0.0047  & 0.032  & 0.048  &  0.075 &  -     & 0.079 & 0.056 / 0.99  && 0.0070  & 0.033 & 0.050  &   -    &    -  & 0.082 &  0.085 / 0.97  \\
\hline\hline
\end{tabular}
\end{center}
\end{table*}